\documentclass[prd,aps,nofootinbib,preprintnumbers,10pt]{revtex4}
\usepackage{amsmath,amssymb,graphicx,epsfig,hyperref,breakurl,color}
\usepackage{lipsum}

\begin{document}

\title{Up-down Asymmetries and Angular Distributions in $D\to K_{1}(\to K\pi\pi)\ell^+\nu_{\ell}$}

\author{Lingzhu Bian~$^1$, Liang Sun~$^1$,  Wei Wang~$^2$ 
}  

\affiliation{ 
$^1$ School of Physics and Technology, Wuhan University, Wuhan, China \\
$^2$ INPAC, Key Laboratory for Particle Astrophysics and Cosmology (MOE),  Shanghai Key Laboratory for Particle Physics and Cosmology, School of Physics and Astronomy, Shanghai Jiao Tong University, Shanghai 200240, China
}

\begin{abstract}
Using the helicity amplitude technique, we derive differential decay widths and  angular distributions for the decay cascade $D\to K_{1}(1270,1400)\ell^+\nu_{\ell}\to (K\pi\pi)\ell^+\nu_{\ell} (\ell=e,\mu)$, in which the electron and muon mass is explicitly included. Using a set of phenomenological results for $D\to K_1$ form factors, we calculate partial  decay widths and branching fractions for $D^0\to K_1^-\ell^+\nu_{\ell}$ and $D^+\to K_1^0\ell^+\nu_{\ell}$, but find that  results for ${\cal B}(D\to K_1(1270)e^+\nu_{e})$ are  larger than recent BESIII measurements by about a factor 1.5.  We further demonstrate that the measurement of up-down asymmetry in  $D\to K_{1}e^+\nu_e\to (K\pi\pi)e^+\nu_{e}$ and angular distributions in  $D\to K_{1}\ell^+\nu_\ell\to (K\pi\pi)\ell^+\nu_{\ell}$ can help  to   determine the hadronic amplitude requested in $B\to K_1(\to K\pi\pi)\gamma$. Based on the Monte-Carlo simulation  with the LHCb geometrical acceptance, we find  that the angular distributions of MC events can be well described. 
\end{abstract}

\maketitle

\section{Introduction}

Since the establishment of standard model  (SM) in 1960s,  searching for new physics (NP) beyond SM has become a most primary objective in particle physics. This can in principle proceed in two distinct directions. It is likely that new particles  emerge directly  in high energy collisions for instance at large hadron collider (LHC).  On the other side,    NP particles can affect  various low-energy  observables by modifying the coupling strength or introducing new interaction forms and thus a  high precision study of these observables  is likely to  indirectly access the NP.
In the SM, the charged weak interaction has the $V-A$ chirality  and thereby
the photon in $b\to s\gamma$ is  predominantly left-handed. The contribution with right-handed polarization is suppressed by the ratio of strange and bottom quark masses.  Therefore the measurement of photon polarization in $b\to s\gamma$  provides a unique probe for  new physics~\cite{Atwood:1997zr,Becirevic:2012dx,Paul:2016urs}. A representative  scenario of this type is the left-right symmetric model~\cite{Kou:2013gna,Haba:2015gwa},  in which the photon  can acquire a significant right-handed component.

In practice, the chirality of the $b\to s\gamma$ can be detected  using the measurements of inclusive $B\to X_s\gamma$ decay branching fractions~\cite{Aubert:2007my,Lees:2012ym,Lees:2012wg,Saito:2014das},  the mixing-induced $CP$ asymmetries of radiative $B^0$ and $B_s^0$ decays~\cite{Ushiroda:2006fi,Aubert:2008gy,Aaij:2019pnd,Akar:2018zhv} and the $B\to K^*e^+e^-$ with very low dilepton mass squared~\cite{Grossman:2000rk,Aaij:2020umj}.  Interestingly, the photon helicity in radiative $D$ decays was also explored~\cite{deBoer:2018zhz}.

In addition to the above methods, it is pointed out that the photon helicity  in $b\to s\gamma$ is proportional to an up-down asymmetry $\mathcal{A}_{\rm UD}$ in $B\to K_1(\to K\pi\pi)\gamma$~\cite{Gronau:2001ng,Gronau:2002rz,Kou:2010kn} and more generally the angular distribution in $B\to K_{res}(\to K\pi\pi)\gamma$. Throughout this work we will use $K_1$ to abbreviate the axial-vector meson $K_1(1270)$ and/or $K_1(1400)$.  However the measurement of up-down asymmetry in $B\to K_1 \gamma$~\cite{Aaij:2014wgo} alone was incapable to reveal the photon helicity  due to the entanglement with the $K_1$ decay dynamics. Many interesting  theoretical analyses have adopted nonperturbative approaches to parametrize the $K_1\to K\pi\pi$  decay amplitude and power constraints on the decay parameters  were obtained~\cite{Gronau:2001ng,Gronau:2002rz,Kou:2010kn,Tayduganov:2011ui,Gronau:2017kyq}.  
In a  previous work~\cite{Wang:2019wee} it is proposed that one can tackle this problem  by combining  semileptonic $D\to  K_1 e^+\nu_{e}$ decays.  In particular,  a ratio of up-down asymmetries in   $D\to K_1(\to K\pi\pi) e^+\nu_e$, ${\cal A}_{UD}^{\prime}$, has been proposed to quantify the hadronic effects in $K_1\to K\pi\pi$ decay. More explicitly  the photon helicity can be expressed as a ratio of the two observables $\lambda_\gamma =3/4 \times   {\cal A}_{UD}/\mathcal{A}_{\rm UD}^{\prime}$~\cite{Wang:2019wee}.

The purpose of this work is multifold. We will first give the details in the helicity amplitude approach to derive the  pertinent angular distributions and up-down asymmetries. Secondly we will extend the previous  analysis to the muon mode whose mass can not be neglected in $D$ decays. Using the phenomenological results for $D\to K_1$ form factors, we calculate partial  decay widths for $D\to K_1\ell^+\nu_{\ell}$, and  show that the measurement of up-down asymmetry in  $D\to K_{1}(\to K\pi\pi) e^+\nu_{e}$ and  the angular distribution in $D\to K_{1}(\to K\pi\pi)\ell^+\nu_{\ell}$ can help  to determine the hadronic amplitude requested in $B\to K_1(\to K\pi\pi)\gamma$. Based on the Monte Carlo (MC) simulation with the LHCb geometrical acceptance, we find  that  the angular distributions of MC events can be well described. 

The rest of this paper is organized as follows. In Sec.~\ref{sec:framework}, we will give a detailed derivation of the angular distributions. In Sec.~\ref{sec:numerics}, we will use the $D\to K_1$ form factors and calculate the differential decay widths. A comparison of predicted branching fractions with BESIII measurements  is made, and a MC simulation of angular distributions with the LHCb geometrical acceptance is also presented.    The last section contains  a brief summary.  

\section{Framework and Angular Distributions}
\label{sec:framework}

\begin{figure}
\centering
\includegraphics[scale=0.7]{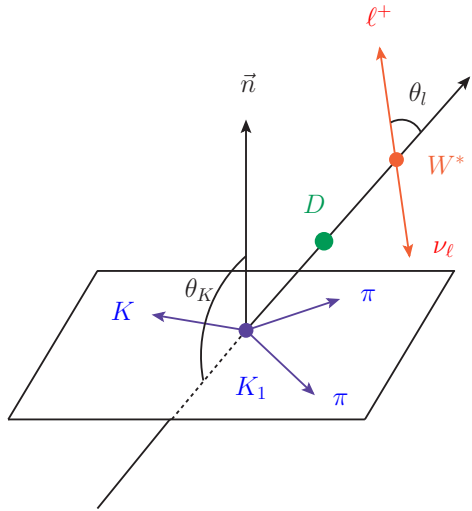}
\caption{The kinematics for $D\to K_{1}\ell^+\nu_{\ell} \to (K\pi\pi)\ell^+\nu_{\ell}$. In the lepton pair $\ell^+\nu_{\ell}$ rest frame, the angle $\theta_\ell$ is defined by the  $\ell^+$  flight direction and the opposite of the $D$ meson flight direction.  In the $K_1$ rest frame the  $\vec n$  is defined as the normal direction of the $K_1$ decay plane, and $\theta_K$ is the relative angle between $\vec n$ and the opposite of the  $D$ flight direction. 
Under parity transformation, the  $D$  flight direction will be reversed, but
since $\vec n$ is a cross product of two momenta, its direction is unchanged. Accordingly the  $\theta_K$ will be changed to $\pi-\theta_K$ under parity transformation, implying that the $\cos\theta_K$ is parity odd.  }
\label{fig:kinematics}
\end{figure}

In this section we will make use of the helicity amplitude technique and derive the angular distributions for the decay cascade $D\to K_1 \ell^+\nu_{\ell} \to (K\pi\pi) \ell^+\nu_{\ell}$. Here the $D$ and $K_1$ could be charged or neutral. Since a neutral $\pi^0$ is   difficult to reconstruct especially at hadron colliders, it is more plausible to  explore the $\pi^+\pi^-$ final state.  Thus we will  mainly consider the decay chain $D^0\to K_1^- \ell^+\nu \to (K^-\pi^+\pi^-) \ell^+\nu_{\ell}$ and $D^+\to \overline  K_1^0 \ell^+\nu \to (\overline K^0\pi^+\pi^-) \ell^+\nu_{\ell}$, though the results are also applicable to other decay channels with neutral pions.  The kinematics of this decay cascade is shown in Fig.~\ref{fig:kinematics}.  In the lepton pair $\ell^+\nu_{\ell}$ rest frame, $\theta_\ell$ is defined by the $\ell^+$ flight direction  and the opposite of the  $D$ meson flight direction.  In $K_1$ rest frame, $\vec n$  is defined as the normal direction of the decay plane, and $\theta_K$ is the relative angle between $\vec n$ and  the opposite of the   $D$ meson flight direction.

A few remarks on the kinematics are given in order. 
\begin{itemize}
\item The normal direction is not unambiguous. For instance, in $K_1^-$ decay plane, it is likely to construct the normal direction with the momentum of $\pi^+$ and $\pi^-$, while the LHCb measurement of up-down asymmetry and angular distributions in  $B\to K_1\gamma$ makes use of the slow and fast pion momentum~\cite{Aaij:2014wgo}, $\vec n\sim \vec p_{\pi, {\rm slow}}\times \vec p_{\pi, {\rm fast}}$. 
\item Secondly, under parity transformation, the  $D$  flight direction will be reversed, but
since $\vec n$ is a cross product of two momenta, its direction is unchanged. Accordingly the  $\theta_K$ will be changed to $\pi-\theta_K$ under parity transformation, implying that the $\cos\theta_K$ is parity odd. The left-handed and right-handed polarization of $K_1$ gives opposite contributions to the $\cos\theta_K$ term. 
\item  Thirdly,   weak interaction in $W^*\to \ell^+\nu_{\ell}$ violates parity conservation. Thus even though $\cos\theta_\ell$ is parity-even, the left-handed and right-handed contributions to  the $\cos\theta_\ell$ term also differ in sign.  
\item Furthermore the  definition of $\theta_K$ depends on   charge or flavor of $K_1$, namely  the angle $\theta_K$  defined in $K_1^-$ decay may differ with the one defined in $K_1^+$ system. It is important to stick with  the same convention on the kinematics in analyzing   $B$ and $D$ decays.  
\end{itemize}

Semileptonic decays of  $D$ into $K_1$ are induced by   effective electro-weak Hamiltonian:
\begin{eqnarray}
{\cal H} = \frac{G_F}{\sqrt{2}}V_{cs} \bar s \gamma^\mu(1-\gamma_5) c \times \bar\nu_{\ell} \gamma_\mu(1-\gamma_5) \ell,
\end{eqnarray} 
where $G_F$ is   Fermi constant, and $V_{cs}$ is   CKM matrix element. With the above Hamiltonian, 
the partial decay width for semileptonic $D$ decays can be  generically  written as 
\begin{align}
d\Gamma =&  \frac{(2\pi)^4}{2m_D} \times d\Phi_n \times  \sum_{spin} {|{\cal M}|^2}. 
\end{align} 
Here $d\Phi_n$ denotes the $n$-body phase space. 
The pertinent decay amplitude  ${\cal M}$ can be decomposed into three individuals: $D\to K_1 W^*$, $K_1\to K\pi\pi$ and $W^*\to \ell^+\nu_{\ell}$. Using the relation between  $g_{\mu\nu}$ and polarization vector,
\begin{align}
  g_{\mu\nu}=-\sum_{\lambda=0,\pm1}\epsilon_\mu^*(\lambda) \epsilon_\nu(\lambda) + \frac{q_\mu q_\nu}{q^2},
\end{align}
one can disassemble decay amplitudes into a hadronic part and leptonic part
\begin{align}
{\cal M} &=\frac{G_F}{\sqrt{2}}V_{cs} H^{\mu}L^{\nu}g_{\mu\nu}  \nonumber\\
&= \frac{G_F}{\sqrt{2}}V_{cs} \left[ -\sum_{\lambda}H\cdot\epsilon^*(\lambda) \times L\cdot\epsilon(\lambda) + H\cdot\epsilon^*(t)\times L\cdot\epsilon(t) \right],
\end{align}
with $\epsilon^{\mu}(t)\equiv q^{\mu}/\sqrt{q^2}$. After the decomposition, both  hadronic   and leptonic parts are Lorentz invariant and thus can be calculated in convenient reference frames. Actually the hadronic part could be further resolved into two individuals, namely $D\to K_1 W^*$ and $K_1\to K\pi\pi$.  Each individuals will be calculated in   rest frame of the decaying parent particle.

\subsection{The leptonic amplitude for $W^*\to \ell^+\nu_{\ell}$}

For simplicity,  we introduce the abbreviation, 
\begin{align}
L(\lambda_e, \lambda_\nu, \lambda_W=0,\pm1) &= L^{\mu}\epsilon_{\mu}(\lambda)= \bar{u}_{\nu}\gamma^{\mu}(1-\gamma_{5})v_{\ell} \epsilon_{\mu}(\lambda), \nonumber\\
L(\lambda_e, \lambda_\nu, \lambda_W=t) &= L^{\mu}\frac{q_\mu}{\sqrt {q^2}} = \bar{u}_{\nu}\gamma^{\mu}(1-\gamma_{5})v_{\ell} \frac{q_\mu}{\sqrt{q^2}},  
\end{align} 
where all spin/helicity indices are explicitly shown. 
Introducing  $f_l= i\sqrt{2(q^2-m_\ell^2)}$ and using  $\hat m_{\ell}= m_{\ell}/\sqrt{q^2}$, one can obtain the  non-vanishing leptonic decay amplitude: 
\begin{eqnarray}
&&L\left(\lambda_e=-\frac{1}{2}, \lambda_\nu=-\frac{1}{2}, \lambda_W=-1\right) = f_l {  \hat m_\ell\sin\theta_\ell} ,\;\;\;
L\left(\lambda_e=-\frac{1}{2}, \lambda_\nu=-\frac{1}{2}, \lambda_W=0\right) =  -f_l{\sqrt2 \hat m_\ell\cos\theta_\ell}, \nonumber\\
&&L\left(\lambda_e=-\frac{1}{2}, \lambda_\nu=-\frac{1}{2}, \lambda_W=1\right) = - f_l{ \hat m_\ell\sin\theta_\ell}, \;\;\;  L\left(\lambda_e=-\frac{1}{2}, \lambda_\nu=-\frac{1}{2}, t\right) =f_l{\sqrt{2}\hat m_\ell} \nonumber\\
&&L\left(\lambda_e=\frac{1}{2}, \lambda_\nu=-\frac{1}{2}, \lambda_W=-1\right) = -f_l (1+\cos\theta_\ell), \;\;\;
L\left(\lambda_e=\frac{1}{2}, \lambda_\nu=-\frac{1}{2}, \lambda_W=0\right) = -f_l\sqrt{2}\sin\theta_\ell, \nonumber\\
&&L\left(\lambda_e=\frac{1}{2}, \lambda_\nu=-\frac{1}{2}, \lambda_W=1\right) =- f_l(1-\cos\theta_\ell). 
\end{eqnarray}
In the massless limit $m_{\ell}\to 0$, only the last three terms are non-zero due to the helicity conservation.

\subsection{$D\to K_1 W^*$}
The $D\to K_1$ transition matrix element is parameterized by a set of form factors: 
\begin{align}
\langle K_1|\bar s \gamma^\mu \gamma_5 c|D \rangle &= -\frac{2iA(q^2)}{m_D-m_{K_1}} \epsilon^{\mu\nu\rho\sigma}(\epsilon^*_{K_1})_\nu (p_D)_\rho (p_{K_1})_\sigma,\\
\langle K_1|\bar s \gamma^\mu c|D \rangle &= -2m_{K_1}V_0(q^2) \frac{\epsilon^*_{K_1}\cdot q}{q^2}q^\mu - (m_D-m_{K_1})V_1(q^2)[\epsilon^*_{K_1}-\frac{\epsilon^{*\mu}_{K_1}\cdot q}{q^2}q^\mu] \nonumber\\
&+ V_2(q^2)\frac{\epsilon^*_{K_1}\cdot q}{m_D-m_{K_1}}[(p_D+p_{K_1})^\mu - \frac{m^2_D-m^2_{K_1}}{q^2}q^\mu], 
\end{align} 
where $q^\mu = p_D^\mu - p_{K_1}^\mu$ is the momentum transfer and $\epsilon^{\mu\nu\rho\sigma}$ is the anti-symmetric Levi-Civita tensor. 
So the matrix element  $c_{\lambda_W}\equiv \langle K_1|\bar s \gamma^\mu(1-\gamma_5) c |D \rangle \epsilon^*_{\mu}(\lambda_{W})$ is evaluated as 
\begin{align}
c_\pm&=(m_D-m_{K_1})V_1\mp\frac{A\sqrt{ \lambda (m^2_D,m^2_{K_1},q^2)}}{m_D-m_{K_1}},\\ 
c_0&=\frac{-1}{2 m_{K_1}  \sqrt {q^2}}\left[(m^2_D-m^2_{K_1}-q^2)(m_D-m_{K_1})V_1-\frac{\lambda (m^2_D,m^2_{K_1},q^2)}{m_D-m_{K_1}}V_2\right],\\
c_t&=-\frac{\sqrt {\lambda (m^2_D,m^2_{K_1},q^2)}}{\sqrt {q^2}}V_0. 
\end{align}
In the above, $\lambda (m^2_D,m^2_{K_1},q^2)= (m^2_D+m^2_{K_1}-q^2)^2-4m^2_Dm^2_{K_1}$.

\subsection{Differential decay width for $D\to K_1\ell^+\nu_{\ell}$}

In this subsection, we will derive  the differential decay width for  $D\to K_1\ell^+\nu_{\ell}$, which serves as  a normalization for the angular distributions for $D\to K_1(\to K\pi\pi)\ell^+\nu_{\ell}$ in the narrow width limit $\Gamma_{K_1}\to 0$.

Combining the three-body phase space
\begin{eqnarray}
d\Phi_3(p_{K_1},p_\ell,p_{\nu_\ell})   &=& \delta^4(p_{D}-p_{K_1}-p_{\ell^+}- p_{\nu_{\ell}})\frac{d^3p_{K_1}}{(2\pi)^3 2E_{K_1}}  \frac{d^3p_{\ell^+}}{(2\pi)^3 2E_{\ell^+}}\frac{d^3p_{\nu_{\ell}}}{(2\pi)^3 2E_{\nu_{\ell}}}\nonumber\\&=& \delta^4(p_{D}-p_{K_1}-q) (2\pi)^4 \delta^4 (q- p_{\ell^+}- p_{\nu_{\ell}}) \frac{d^4q}{(2\pi)^4} \frac{d^3p_{K_1}}{(2\pi)^3 2E_{K_1}}  \frac{d^3p_{\ell^+}}{(2\pi)^3 2E_{\ell^+}}\frac{d^3p_{\nu_{\ell}}}{(2\pi)^3 2E_{\nu_{\ell}}}\nonumber\\
&=& \frac{1}{(2\pi)^7}\frac{\sqrt{\lambda(m^2_D, m^2_{K_1}, q^2)}  }{32m_{D}^2 }(1-m_{\ell}^2/q^2)  \times  d\cos \theta_\ell  dq^2, 
\end{eqnarray} 
one can obtain the angular distribution for $D\to K_1\ell^+\nu_{\ell}$
\begin{align}
\frac{d\Gamma}{dq^2 d\cos\theta_{\ell}} 
=&   \frac{ G^2_F |V_{cs}|^2\sqrt{\lambda(m^2_D, m^2_{K_1}, q^2)}q^2  }{512\pi^3 m_{D}^3 }(1-\hat m_{\ell}^2)^2  \times    \bigg( 2c_0^2({\sin}^2 \theta_\ell + \hat m_\ell^2{\cos}^2 \theta_\ell)\nonumber\\
+& c_+^2[(1+\cos \theta_\ell)^2 + \hat m_\ell^2\sin^2 \theta_\ell ]+c_-^2[(1-\cos \theta_\ell)^2+\hat m_\ell^2\sin^2 \theta_\ell ] + c^2_t  2\hat m^2_l  -{\rm  Re}[c_0c^*_t] {4\hat m^2_l\cos \theta} \bigg). 
\end{align}
Integrating over $\cos\theta_\ell$, one can have 
 partial decay widths~\cite{Li:2009tx,Wang:2015cis}: 
\begin{eqnarray}
\frac{d\Gamma_L(D\to K_1\ell^+\nu_{\ell})}{dq^2}&=&   \frac{ G^2_F |V_{cs}|^2\sqrt{\lambda(m^2_D, m^2_{K_1}, q^2)}q^2  }{512\pi^3 m_{D}^3 }(1-\hat m_{\ell}^2)^2  \times  \left(\frac{4}{3} c_0^2 (2+\hat m_{\ell}^2)+ 4 \hat m_{\ell}^2 c_t^2\right)  \nonumber\\
&=&  
 \frac{\sqrt {\lambda(m^2_D,m^2_{K_1},q^2)}G^2_F|V_{cs}|^2}{384m^3_D\pi^3} (1-\hat m_{\ell}^2)^2  \times  \bigg\{3\hat m^2_l\lambda(m^2_D,m^2_{K_1},q^2)V^2_0 \nonumber\\
&&+  (\hat m^2_\ell+2 )|\frac{1}{2m_{K_1}}[(m^2_D-m^2_{K_1}-q^2)(m_D-m_{K_1})V_1-\frac{\lambda(m^2_D,m^2_{K_1},q^2)}{m_D-m_{K_1}}V_2]|^2\bigg\}, 
\end{eqnarray}
\begin{eqnarray}
\frac{d\Gamma_{\pm}(D\to K_1\ell^+\nu_{\ell})}{dq^2}&=&    \frac{ G^2_F |V_{cs}|^2\sqrt{\lambda(m^2_D, m^2_{K_1}, q^2)}q^2  }{512\pi^3 m_{D}^3 }(1-\hat m_{\ell}^2)^2  \times  \frac{4}{3} c_\pm^2 (2+\hat m_{\ell}^2) \nonumber\\
&=&   \frac{\sqrt {\lambda(m^2_D,m^2_{K_1},q^2)}G^2_F|V_{cs}|^2}{384m^3_D\pi^3}(1-\hat m_{\ell}^2)^2 q^2  \nonumber \\
&&\times  \bigg\{(\hat m^2_\ell+2)\lambda(m^2_D,m^2_{K_1},q^2)|\frac{A}{m_D-m_{K_1}}\mp \frac{(m_D-m_{K_1})V_1}{\sqrt {\lambda (m^2_D,m^2_{K_1},q^2)}}|^2\bigg\}, 
\end{eqnarray}
where $L$ and $\pm$ in the subscripts denote contribution from longitudinal and transverse polarization.

\subsection{$K_1\to K\pi\pi$}

The hadronic part in the decay cascade $D\to K_1 \ell^+\nu_{\ell} \to (K\pi\pi) \ell^+\nu_{\ell}$ contains: 
\begin{align}
&H\cdot \epsilon^{*}_{W}(\lambda)\sim \langle   K\pi\pi|K_1\rangle  \times \langle K_1|(V-A)_\mu|D \rangle \epsilon^{*\mu}_{W}(\lambda),
\end{align}
where $\langle   K\pi\pi|K_1\rangle$ is parameterized as
\begin{align}
\langle   K\pi\pi| K_1\rangle = (2\pi)^4 \delta^4(p_{K_1}- p_{K}-p_\pi-p_\pi) \times \epsilon_{K_1} \cdot   J. 
\end{align} 
Notice that the explicit form of $J$ depends on the convention of the $\vec n$.   

In the  $K_1$ rest frame, one can set the normal direction as the $z$-axis and the momenta of $(K,\pi^+, \pi^-)$ lies in the $x-y$ plane.  Since $J$ is a linear combination of the momenta of two pions,  $J_z=0$ and 
\begin{align}
 J_\mu= (J_0,J_x,J_y,0). 
\end{align}
To simplify the calculation, we choose $K_1$  moving along $(\theta_{K}, \phi)$ direction, and thus $\epsilon_{K_1} \cdot   J$ is evaluated as 
\begin{align}
&\epsilon_{K_1}(0) \cdot   J = \sin \theta_K(J_x\cos \phi+J_y\sin \phi),\\ 
&\epsilon_{K_1}(1) \cdot   J = -\frac{1}{\sqrt 2}[\cos \phi(J_x\cos \theta_K+iJ_y)+\sin \phi(J_y\cos \theta_K-iJ_x)],\\
&\epsilon_{K_1}(-1)\cdot   J=\frac{1}{\sqrt 2}[\cos \phi(J_x\cos \theta_K-iJ_y)+\sin \phi(J_y\cos \theta_K+iJ_x)]. 
\end{align}

\subsection{Angular Distributions in $D\to  K_1\ell^+\nu_{\ell}\to  (K\pi\pi) \ell^+\nu_{\ell}$}

With the above individuals, one   obtains the total decay amplitude: 
\begin{align}
{\cal M}(\lambda_\ell=-\frac{1}{2})&=L\left(\lambda_e=-\frac{1}{2}, \lambda_\nu=-\frac{1}{2}, \lambda_W=-1\right) \times \epsilon_{K_1}(-1)\cdot   J \times c_- \nonumber\\
&+L\left(\lambda_\ell=-\frac{1}{2}, \lambda_\nu=-\frac{1}{2}, \lambda_W=1\right) \times \epsilon_{K_1}(1)\cdot   J \times c_+ \nonumber\\
&+L\left(\lambda_\ell=-\frac{1}{2}, \lambda_\nu=-\frac{1}{2}, \lambda_W=0\right) \times \epsilon_{K_1}(0)\cdot   J \times c_0 \nonumber\\
&-L\left(\lambda_\ell=-\frac{1}{2}, \lambda_\nu=-\frac{1}{2}, \lambda_W=t\right) \times \epsilon_{K_1}(0)\cdot   J \times c_t, \\
{\cal M}(\lambda_\ell=\frac{1}{2})&=L\left(\lambda_e=\frac{1}{2}, \lambda_\nu=-\frac{1}{2}, \lambda_W=-1\right) \times \epsilon_{K_1}(-1)\cdot   J \times c_- \nonumber\\
&+L\left(\lambda_\ell=\frac{1}{2}, \lambda_\nu=-\frac{1}{2}, \lambda_W=1\right) \times \epsilon_{K_1}(1)\cdot   J \times c_+ \nonumber\\
&+L\left(\lambda_\ell=\frac{1}{2}, \lambda_\nu=-\frac{1}{2}, \lambda_W=0\right) \times \epsilon_{K_1}(0)\cdot   J \times c_0 \nonumber\\
&-L\left(\lambda_e=\frac{1}{2}, \lambda_\nu=-\frac{1}{2}, \lambda_W=t\right) \times \epsilon_{K_1}(0)\cdot   J \times c_t.
\end{align}
Using two abbreviations:
\begin{align}
|J|^2=|J_x|^2+|J_y|^2, \;\;\;{\rm Im}[n\cdot(\vec J \times \vec J^*)]=-i(J_xJ^*_y-J_yJ^*_x),
\end{align}
and integrating over $\phi$,  we obtain:
\begin{align}
|{\cal M}|^2&=\frac{3}{4\pi|J|^2}\int d\phi (|{\cal M}(\lambda_e=\frac{1}{2})|^2+|{\cal M}(\lambda_e=-\frac{1}{2})|^2)\nonumber\\
&=\frac{3}{8}(d_1+d^\prime_1[\cos^2 \theta_K\cos^2 \theta_\ell] + d_2\cos \theta_\ell + d^\prime_2 \cos^2 \theta_K\cos \theta_\ell \nonumber\\
&+ d_3\cos \theta_K + d^\prime_3\cos \theta_K \cos^2 \theta_\ell + d_4\cos \theta_K\cos \theta_\ell + d_5\cos^2 \theta_K+d^\prime_5\cos^2 \theta_\ell),
\end{align}
where  a factor $\frac{3}{4\pi|J|^2}$ is introduced to be consistent with the three-body decay width. 
In the above equation, the angular coefficients are calculated  as
\begin{align}
&d_1=(1+\hat m^2_\ell)(|c_-|^2+|c_+|^2)+4|c_0|^2+4\hat m^2_\ell|c_t|^2,  \nonumber\\
&d^\prime_1=(1-\hat m^2_\ell)(4|c_0|^2+|c_-|^2+|c_+|^2),  \nonumber\\
&d_2=-2[|c_-|^2-|c_+|^2+4{\rm Re}[c_0c^*_t]\hat m^2_\ell],  \nonumber\\
&d^\prime_2=-2[|c_-|^2-|c_+|^2-4 {\rm Re}[c_0c^*_t]\hat m^2_\ell],  \nonumber\\
&d_3=2\frac{{\rm Im}[\vec n\cdot (\vec J \times \vec J^*)]}{|J|^2}[(1+\hat m^2_\ell)(|c_+|^2-|c_-|^2)] , \nonumber\\
&d^\prime_3=2\frac{{\rm Im}[\vec n\cdot (\vec J \times \vec J^*)]}{|J|^2}[(1-\hat m^2_\ell)(|c_+|^2-|c_-|^2)]  \nonumber\\
&d_4=4\frac{{\rm Im}[\vec n\cdot (\vec J \times \vec J^*)]}{|J|^2}(|c_-|^2+|c_+|^2),  \nonumber\\
&d_5=-[(1+\hat m^2_\ell)(-|c_-|^2-|c_+|^2)+4|c_0|^2+4\hat m^2_\ell|c_t|^2],  \nonumber\\
&d^\prime_5=-[(1-\hat m^2_\ell)(4|c_0|^2-|c_-|^2-|c_+|^2)].
\end{align}
Apparently  the following combination can be used to extract the hadron amplitude 
\begin{eqnarray}
 \frac{d_3+d_3'}{d_2+d_2'} = \frac{{\rm Im}[\vec n\cdot (\vec J \times \vec J^*)]}{|J|^2}. 
 \end{eqnarray}

Including the phase-space, we arrive at the angular distribution for $D\to K_1\ell^+\nu_{\ell}\to  K\pi\pi \ell^+\nu_{\ell}$ as:
\begin{align}
\frac{d\Gamma}{dq^2 d\cos \theta_\ell   d\cos \theta_K}
=&\frac{G^2_F|V_{cs}|^2q^2\sqrt{\lambda(m^2_D, m^2_{K_1}, q^2)}  }{512\pi^3 m_{D}^3 }(1-\hat m_{\ell}^2)^2 \nonumber  \\
\times&\frac{3}{8}\bigg(d_1+d^\prime_1[\cos^2 \theta_K\cos^2 \theta_\ell] + d_2\cos \theta_\ell + d^\prime_2 \cos^2 \theta_K\cos \theta_\ell\nonumber\\
+& d_3\cos \theta_K + d^\prime_3\cos \theta_K \cos^2 \theta_\ell + d_4\cos \theta_K\cos \theta_\ell + d_5\cos^2 \theta_K+d^\prime_5\cos^2 \theta_\ell\bigg). 
\end{align}
The ratio of differential  up-down asymmetries proposed in Ref.~\cite{Wang:2019wee} is evaluated as:
\begin{eqnarray}\label{eq:AUD2}
{\cal A}_{\rm UD}^{\prime}  & \equiv& \frac{\frac{d\Gamma}{dq^2}[\cos\theta_K>0]-\frac{d\Gamma}{dq^2}[\cos\theta_K<0]}{ \frac{d\Gamma}{dq^2}[\cos\theta_\ell>0]-\frac{d\Gamma}{dq^2}[\cos\theta_\ell<0]} \nonumber\\
&=& \frac{3d_3+d_3'}{3d_2+d_2'} \nonumber\\
&=&  \frac{{\rm Im}[\vec n\cdot (\vec J \times \vec J^*)]}{|J|^2}\frac{(2+\hat m_{\ell}^2) (|c_-|^2-|c_+|^2)}{2[(|c_-|^2-|c_+|^2)+2{\rm Re}[c_0c_t^*] \hat m_{\ell}^2]}. \label{eq:differential_UD}
\end{eqnarray}
If the massless limit $\hat m_{\ell}\to 0$, the above ratio is reduced to the hadronic amplitude $\frac{{\rm Im}[\vec n\cdot (\vec J \times \vec J^*)]}{|J|^2}$, but apparently this reduction is contaminated by  the lepton mass. 

\subsection{Angular Distributions in $B\to K_1\gamma$}

This subsection gives   the angular distribution in $B\to K_{1}(\to K\pi\pi)\gamma$. The effective Hamiltonian for $b\to s\gamma$ has the general form: 
\begin{align}
{\cal H}_{{\rm eff}} & =-\frac{4G_{F}}{\sqrt{2}}V_{tb}V_{ts}^{*}(C_{7L}{\cal O}_{7L}+C_{7R}{\cal O}_{7R}),\nonumber \\
 & \qquad{\cal O}_{7L,R}=\frac{em_{b}}{16\pi^{2}}\bar{s}\sigma_{\mu\nu}\frac{1\pm\gamma_{5}}{2}bF^{\mu\nu},
\end{align}
where $C_{7L,7R}$ are  the corresponding Wilson coefficients  for ${\cal O}_{7L,R}$, and $V_{tb},V_{ts}$ are CKM matrix elements.
Due to the chirality  structure of $W^{\pm}$ in  
SM,  the   photon in $b\to s\gamma$ is predominantly  left-handed, while the right-handed polarization is suppressed by approximately $ m_{s}/m_{b}$.

Using the helicity amplitude technique one can similarly calculate the angular distributions for  $ {B}\to K_{1}(\to {K}\pi\pi)\gamma$, and the results are   easier in two aspects. First,  there is no leptonic part in the decay cascade. 
Secondly, the $B\to K_1 \gamma$ decay amplitudes only contain two polarizations. Without including higher order QCD corrections,  these two polarizations are proportional  to $C_{7L,7R}$. So $ {B}\to K_{1}(\to {K}\pi\pi)\gamma$
has the  differential decay rate~\cite{Gronau:2001ng,Gronau:2002rz,Gronau:2017kyq}:  
\begin{eqnarray}\label{eq:BK1g}
&&\frac{d\Gamma_{ K_1\gamma}}{d\cos\theta_K}=\frac{|A|^2|\vec{J}|^{2}}{4} \times   \left[1+\cos^{2}\theta_K+2\lambda_{\gamma}\cos\theta_K\frac{{\rm Im}[\vec{n}\cdot(\vec{J}\times\vec{J}^{*})]}{|\vec{J}|^{2}}\right]. 
\end{eqnarray}
In this equation, the nonperturbative amplitude $A$  characterizes the $B\to K_1\gamma$, and the $\theta_K$ is the same angle as in Fig.~\ref{fig:kinematics}.  
The  photon helicity $\lambda_{\gamma}$ is
\begin{equation}\label{eq:lambdagamma}
\lambda_{\gamma}\equiv\frac{|\mathcal{A}(  B\to   K_{1R}\gamma_R)|^{2}-|\mathcal{A}(  B\to   K_{1L}\gamma_L)|^{2}}{|\mathcal{A}(  B\to   K_{1R}\gamma_R)|^{2}+|\mathcal{A}(  B\to   K_{1L}\gamma_L)|^{2}}, 
\end{equation}
with  $\lambda_{\gamma}\simeq-1$ for $b\to s\gamma$ but $\lambda_{\gamma}\simeq+1$ for $\bar b\to \bar s\gamma$    in SM.

\section{Numerical Results and Discussions }
\label{sec:numerics}

\subsection{$K_1$ mixing and $D\to K_1$ form factors}

\begin{table}
\caption{The $D\to K_1$ form factors calculated in the covariant  LFQM~\cite{Cheng:2003sm}.  The physical $K_1$ states ($K_1(1270)$ and $K_1(1420)$) are mixtures of the $K_{1A}$ ($^3P_1$) and $K_{1B}$  ($^1P_1$).  }
 \label{Tab:formFactor_D_K1}
 \begin{center}
 \begin{tabular}{|c|c|c|c|c|c|c|c|}
\hline \hline
         $F$   &$F(0)$   &$a$   &$b$ &$F$   &$F(0)$   &$a$   &$b$               \\
\hline
$A^{D K_{1A}}$
    & $0.15^{-0.01+0.01}_{+0.01-0.01}$
    & $0.89^{-0.03+0.00}_{+0.03-0.01}$
    & $0.12^{-0.02-0.01}_{+0.02+0.01}$
    & $V^{D K_{1A}}_0$
    & $0.28^{-0.00-0.00}_{+0.00+0.00}$
    & $0.84^{-0.02-0.01}_{+0.01-0.01}$
    & $0.39^{-0.05+0.04}_{+0.06-0.03}$
    \\
$V^{D K_{1A}}_1$
    & $1.60^{-0.05-0.02}_{+0.05+0.01}$
    & $-0.22^{-0.00-0.03}_{+0.00+0.03}$
    & $0.07^{-0.00+0.00}_{+0.00-0.00}$
    & $V^{D K_{1A}}_2$
    & $0.01^{-0.00+0.00}_{+0.00-0.00}$
    & $-0.83^{-0.17+0.02}_{+0.15-0.03}$
    & $0.24^{+0.04-0.01}_{-0.03+0.01}$
    \\
\hline    
$A^{D K_{1B}}$
    & $0.10^{+0.00+0.00}_{-0.00-0.00}$
    & $0.98^{-0.01-0.01}_{+0.01+0.01}$
    & $0.37^{-0.03-0.03}_{+0.03+0.04}$
    & $V^{D K_{1B}}_0$
    & $0.48^{-0.01+0.02}_{+0.01-0.03}$
    & $0.94^{-0.02-0.02}_{+0.01+0.01}$
    & $0.22^{+0.00-0.03}_{-0.00+0.03}$
    \\
$V^{D K_{1B}}_1$
    & $1.58^{+0.02+0.03}_{-0.03-0.05}$
    & $0.31^{-0.02-0.01}_{+0.02+0.02}$
    & $0.04^{-0.00-0.00}_{+0.00+0.01}$
    & $V^{D K_{1B}}_2$
    & $-0.13^{+0.01-0.01}_{-0.01+0.01}$
    & $0.57^{-0.06-0.01}_{+0.04-0.01}$
    & $0.32^{+0.05-0.04}_{-0.03+0.06}$
    \\ 
\hline
\end{tabular}
\end{center}
\end{table}

Since   strange quark is heavier than up/down quark, $K_1(1270)$ and $K_1(1400)$ are not purely $K_{1A}(^3P_1)$ and $K_{1B}(^1P_1)$ states, and instead they   mix:
\begin{eqnarray}
|K_1(1270)\rangle&=&|K_{1A}\rangle
{\rm{sin}}\Theta_K+|K_{1B}\rangle{\rm{cos}}\Theta_K,\\
|K_1(1400)\rangle&=&|K_{1A}\rangle
{\rm{cos}}\Theta_K-|K_{1B}\rangle{\rm{sin}}\Theta_K.
\end{eqnarray}
The mixing angle $\Theta_K$ can be determined by decays such  as  $\tau^-\to K_1^-\nu_\tau$, whose   decay rate is 
\begin{eqnarray}
{\Gamma}(\tau^-\to
K_1^-\nu_\tau)=\frac{m_\tau^3}{16\pi}G_F^2|V_{us}|^2f_A^2\left(1-\frac{m_A^2}{m_\tau^2}
\right)^2\left(1+\frac{2m_A^2}{m_\tau^2}\right). 
\end{eqnarray}
From the measured branching fractions~\cite{Zyla:2020zbs}  
\begin{eqnarray}
{\cal B}(\tau^-\to K_1(1270)\nu_\tau)=(4.7\pm1.1)\times 10^{-3},\\
{\cal B}(\tau^-\to K_1(1400)\nu_\tau)=(1.7\pm2.6)\times
10^{-3},\label{eq:BRintau}
\end{eqnarray}
one can determine $K_1$ decay constants: 
\begin{eqnarray}
|f_{K_1(1270)}|=(169_{-21}^{+19}){\rm MeV};\;\;\;
|f_{K_1(1400)}|=(125_{-125}^{+~74}){\rm MeV}.\label{eq:decayconstantK1}
\end{eqnarray}
Combing the QCD sum rules for $K_{1A}$ and
$K_{1B}$ decay constants~\cite{Yang:2007zt},  one determines the mixing angle with a fourfold ambiguity~\cite{Li:2009tx}: 
\begin{eqnarray}
 -143^\circ<\Theta_K<-120^\circ,\;\;\;{\rm or}\;\;
 -49^\circ<\Theta_K<-27^\circ,\;\;\;{\rm or}\;\;
 37^\circ<\Theta_K<60^\circ,\;\;\;{\rm or}\;\;
 131^\circ<\Theta_K<153^\circ.
\end{eqnarray}
Except the third scenario, the other scenarios are not favored by $K_1$ masses or $B\to K_1\gamma$ decay widths~\cite{Abe:2004kr}. Actually the $\tau\to K_1\nu$ may give direct information on the $K_1$ decays~\cite{Hayasaka:2021ecj}.  With other available  constraints,  Ref.~\cite{Verma:2011yw} suggested the use of $\Theta_K=50.8^\circ$, while $\Theta_K=33^\circ$ is suggested in Ref.~\cite{Cheng:2017pcq}. In the following we will use $\Theta_K=60^\circ$ as the central result (which is also favored by BESIII measurements of ${\cal B}(D\to K_1e^+\nu_{e})$), but the dependence on $\Theta_K$ in a wider range  $30^\circ<\Theta_K<60^\circ$ will be presented.

The $D\to K_{1A,1B}$ form factors have been calculated in covariant light-front quark model (LFQM)~\cite{Cheng:2003sm,Chang:2020wvs,Verma:2011yw,Cheng:2017pcq}, and the updated results from Ref.~\cite{Verma:2011yw} are collected in Tab.~\ref{Tab:formFactor_D_K1}. In the calculation, 
the $q^2$-distribution of form factors is parametrized as:
\begin{eqnarray}
F(q^2) &=& \frac{F(0)}{1- a q^2/m_D^2+ b(q^2/m_D^2)^2}, 
\end{eqnarray}
but a different parametrization is adopted for $V_2^{D\to K_{1B}}$~\cite{Cheng:2003sm}: 
\begin{eqnarray}
F(q^2) &=& \frac{F(0)}{(1-q^2/m_D^2)(1- a q^2/m_D^2+ b(q^2/m_D^2)^2)}.  
\end{eqnarray}
Physical form factors are obtained through: 
\begin{eqnarray}
F^{D\to K_1(1270)} &=&F^{D\to K_{1A}} 
{\rm{sin}}\Theta_K+F^{D\to K_{1B}}{\rm{cos}}\Theta_K,\\
F^{D\to K_1(1400)} &=&F^{D\to K_{1A}} 
{\rm{cos}}\Theta_K-F^{D\to K_{1B}} {\rm{sin}}\Theta_K.
\end{eqnarray}
The $D\to K_1$ form factors have also been calculated in other approaches~\cite{Momeni:2019uag,Khosravi:2008jw,Momeni:2020zrb} but results differ significantly. 

\subsection{Decay widths and branching fractions}

\begin{figure}
\centering
\includegraphics[scale=0.7]{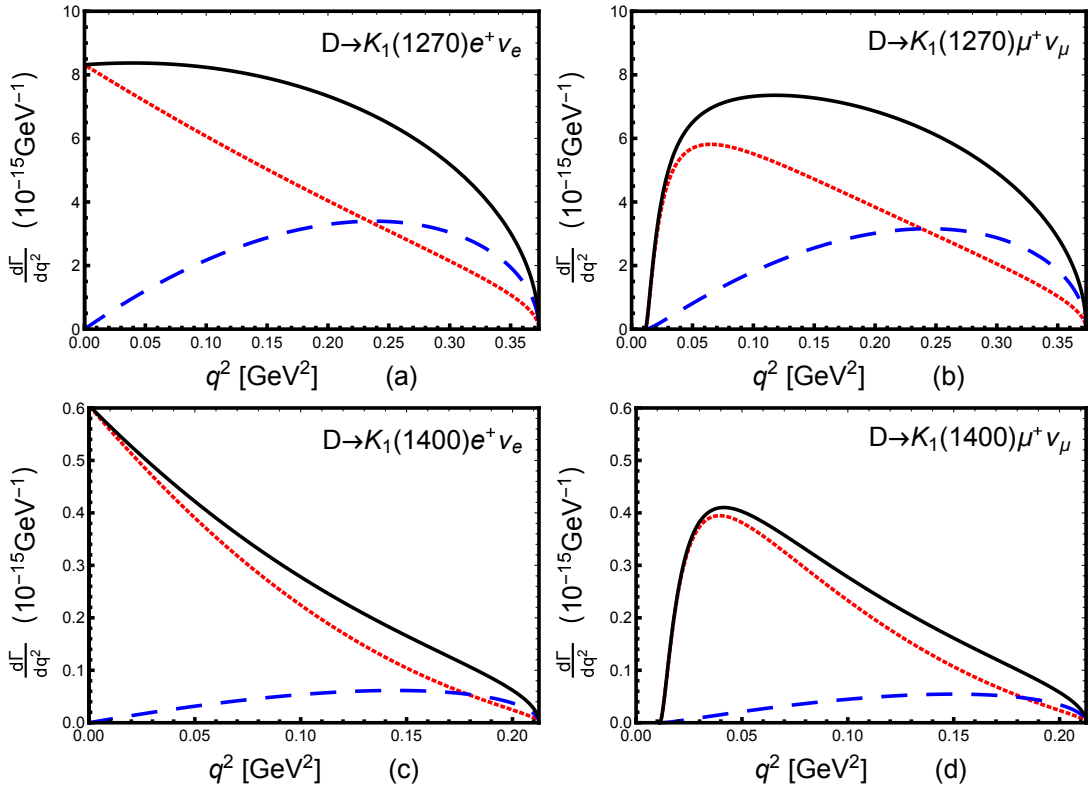}
\caption{Differential decay widths for $D\to K_1(1270)\ell^+\nu_{\ell}$ (in units of $10^{-15}{\rm GeV}^{-1}$) and $D\to K_1(1400)\ell^+\nu_{\ell}$ (in units of $10^{-15}{\rm GeV}^{-1}$). The dotted and dashed lines correspond to the longitudinal and transverse polarizations, while the solid line gives the total differential decay widths.  }
\label{fig:dgammadq2K11270}
\end{figure} 

 \begin{table}
 \caption{Results for integrated branching ratios for the $D\to K_1(1270)\ell^+\nu_{\ell}$ and $D\to K_1(1400)\ell^+\nu_{\ell}$ decays (in  units of $10^{-3}$). The experimental results are taken from BESIII measurements~\cite{Ablikim:2019wxs,Ablikim:2021ocw}.  }
 \label{tab:br_DK_1}
 \begin{center}
 \begin{tabular}{c|cccc|ccccc}
 \hline\hline     $D^0\to K^-(1270)\ell^+\nu_{\ell}$ & ${\cal B}_{\rm{L}}$  &${\cal B}_{\rm T}$ &${\cal B}_{\rm{L}}/{\cal B}_{\rm{T}}$   &${\cal B}_{\rm{total}}$    &${\cal B}_{\rm{data}}$\\\hline $\ell=e$ & 1.01 &  0.56  &  1.80 &   1.56&$1.09 \pm 0.13^{+0.09}_{-0.13}\pm 0.12$\\$\ell=\mu$ & 0.81 &  0.49  &  1.60 &   1.30&--\\\hline     $D^+\to \overline K^0(1270) \ell^+\nu_{\ell}$ & ${\cal B}_{\rm{L}}$  &${\cal B}_{\rm T}$ &${\cal B}_{\rm{L}}/{\cal B}_{\rm{T}}$   &${\cal B}_{\rm{total}}$   &${\cal B}_{\rm{data}}$\\\hline$\ell=e$ & 2.56 &  1.41  &  1.80 &   3.97&$2.30 \pm 0.26^{+0.18}_{-0.21} \pm0.25$\\$\ell=\mu$ & 2.06 &  1.25  &  1.60 &   3.31&--\\\hline     $D^0\to K^-(1400)\ell^+\nu_{\ell}$ & ${\cal B}_{\rm{L}}$  &${\cal B}_{\rm T}$ &${\cal B}_{\rm{L}}/{\cal B}_{\rm{T}}$   &${\cal B}_{\rm{total}}$  \\\hline$\ell=e$ & 0.031 &  0.006  &  5.60 &   0.037&--\\$\ell=\mu$ & 0.024 &  0.005  &  5.30 &   0.029&--\\\hline     $D^+\to \overline K^0(1400) \ell^+\nu_{\ell}$ & ${\cal B}_{\rm{L}}$  &${\cal B}_{\rm T}$ &${\cal B}_{\rm{L}}/{\cal B}_{\rm{T}}$   &${\cal B}_{\rm{total}}$\\\hline   $\ell=e$ & 0.080 &  0.014  &  5.60 &   0.094&--\\$\ell=\mu$ & 0.062 &  0.012  &  5.30 &   0.074&--\\\hline \hline 
 \end{tabular}
 \end{center}
 \end{table}

To calculate decay widths, we use the following inputs from Particle Data Group~\cite{Zyla:2020zbs}:
\begin{eqnarray}
\tau(D^0)=(0.4101\pm0.0015)\times 10^{-12}s ,\;\;\tau(D^+)=(1.040\pm0.007)\times 10^{-12}s,\nonumber\\
m_{K_1(1270)}=1.253{\rm GeV}, \;\;\; m_{K_1(1400)}=1.403{\rm GeV}, \;\;\; V_{cs}=0.973. 
\end{eqnarray}  
Differential decay widths  $d\Gamma/dq^2$ (in units of $10^{-15}{\rm GeV}^{-1}$) for $D\to K_1(1270)\ell^+\nu_{\ell}$ and $D\to K_1(1400)\ell^+\nu_{\ell}$   are shown in Fig.~\ref{fig:dgammadq2K11270}. The dotted and dashed lines correspond to the longitudinal and transverse polarizations, while the solid line gives the total differential decay widths.  At low $q^2$, the longitudinal polarization dominates.

Results for integrated branching ratios for the $D\to K_1(1270)\ell^+\nu_{\ell}$ and $D\to K_1(1400)\ell^+\nu_{\ell}$ decays (in  units of $10^{-3}$) are given in Tab.~\ref{tab:br_DK_1}. The experimental results are taken from BESIII measurements~\cite{Ablikim:2019wxs,Ablikim:2021ocw}. Through this table, one can see that the theoretical results for branching fractions of $D\to K_1(1270)e^+\nu_e$ are  larger than the experimental data by about a factor of 1.5.  However  we find that using  form factors from Refs.~\cite{Momeni:2019uag,Khosravi:2008jw,Momeni:2020zrb} the branching fractions can be significantly reduced.  Branching fractions for $D\to K_1(1400)\ell^+\nu_{\ell}$ are suppressed by orders of magnitudes, and this pattern is consistent with the BESIII observations~\cite{Ablikim:2019wxs,Ablikim:2021ocw}.

\begin{figure}
\centering
\includegraphics[scale=0.7]{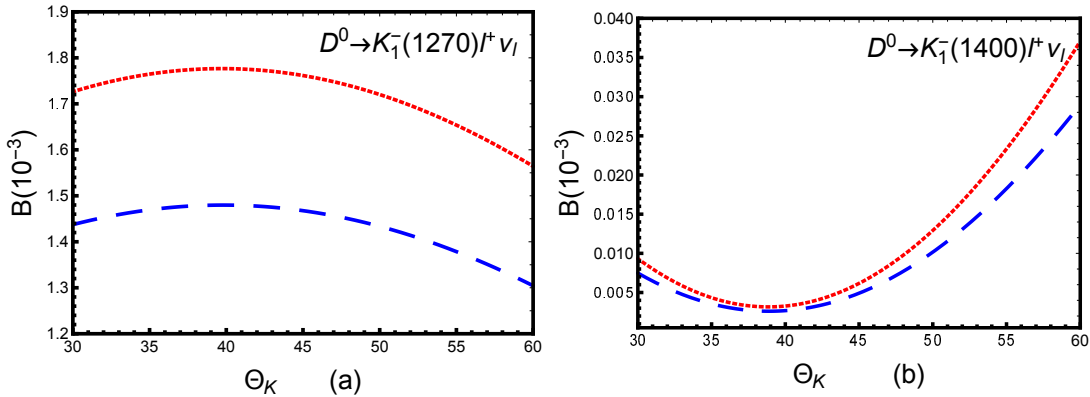}
\caption{Dependence of branching ratios ${\cal B}(D^0\to K_1^-\ell^+\nu_{\ell})$ (in units of $10^{-3}$) on the mixing angle $\Theta_K$ in the range  $30^\circ<\Theta_K<60^\circ$.  Dotted and dashed curves correspond to the electron and muon mode, respectively.    }
\label{fig:BRmixingAngle}
\end{figure}

Fig.~\ref{fig:BRmixingAngle} shows the dependence of branching fractions ${\cal B}(D^0\to K_1^-\ell^+\nu_{\ell})$ (in units of $10^{-3}$) on the mixing angle $\Theta_K$ in the range  $30^\circ<\Theta_K<60^\circ$.  Dotted and dashed curves correspond to the electron and muon mode, respectively. Due to the phase space suppression, decays with muon in the final state are typically smaller than the electron mode by approximately $10\%-20\%$.

\subsection{Angular distributions}

\begin{figure}
\centering
\includegraphics[scale=0.7]{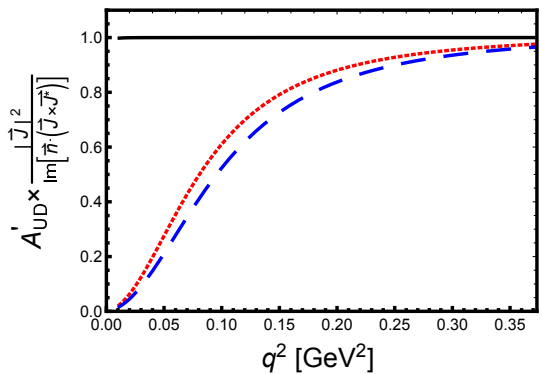}
\caption{Results for up-down asymmetry $A_{UD}'$ (in unit of ${\rm Im}[\vec n\cdot (\vec J\times \vec J^*)]/|\vec J|^2$) in $D\to K_1(\to K\pi\pi)\ell^+\nu_{\ell}$. The solid curve corresponds to  the electron final state, in which  the lepton mass is negligible  and the result is very close to unity. The dashed and dotted results correspond to the muon mode with the mixing angle $\Theta_K=60^\circ$ and $\Theta_K=30^\circ$, respectively.  The nonzero mass of $\mu$ provides  sizable corrections as shown in Eq.~\eqref{eq:differential_UD}.   }
\label{fig:Audplot_theory}
\end{figure} 

Results for up-down asymmetry $A_{UD}'$ (in unit of ${\rm Im}[\vec n\cdot (\vec J\times \vec J^*)]/|\vec J|^2$) in $D\to K_1(1270)\ell^+\nu_{\ell}$ are shown in Fig.~\ref{fig:Audplot_theory}. The solid curve corresponds to  the electron final state, in which  the lepton mass is negligible  and the result is very close to unity. The dashed and dotted results correspond to the muon mode with the mixing angle $\Theta_K=60^\circ$ and $\Theta_K=30^\circ$, respectively. From this figure, one can see that the nonzero mass of muon can give considerable  corrections also shown in Eq.~\eqref{eq:differential_UD}, but the results are less sensitive to the $D\to K_1$ form factors.

Integrating over the $q^2$, we obtain the angular distributions of $D\to K_{1}(\to K\pi\pi)\ell^+\nu_{\ell}$ dataset as
\begin{align}
\frac{d\Gamma}{d\cos \theta_\ell d\cos \theta_K}=&a_1+a_2[\cos^2 \theta_K\cos^2 \theta_\ell] + a_3\cos \theta_\ell + a_4\cos^2 \theta_K\cos \theta_\ell \nonumber\\
+& a_5\cos \theta_K + a_6\cos \theta_K \cos^2 \theta_\ell + a_7\cos \theta_K\cos \theta_\ell + a_8\cos^2 \theta_K+a_9\cos^2 \theta_\ell, \label{con:formula_mu}
\end{align}
with 
\begin{eqnarray}
a_i = \frac{3}{8}\int dq^2 \frac{G^2_FV^2_{cs}q^2\sqrt{\lambda(m^2_D, m^2_{K_1}, q^2)}  }{512\pi^3 m_{D}^3 }(1-m_{\ell}^2/q^2)^2  \times  d_i. 
\end{eqnarray}

\begin{figure}
    \centering
    \includegraphics[scale=0.80]{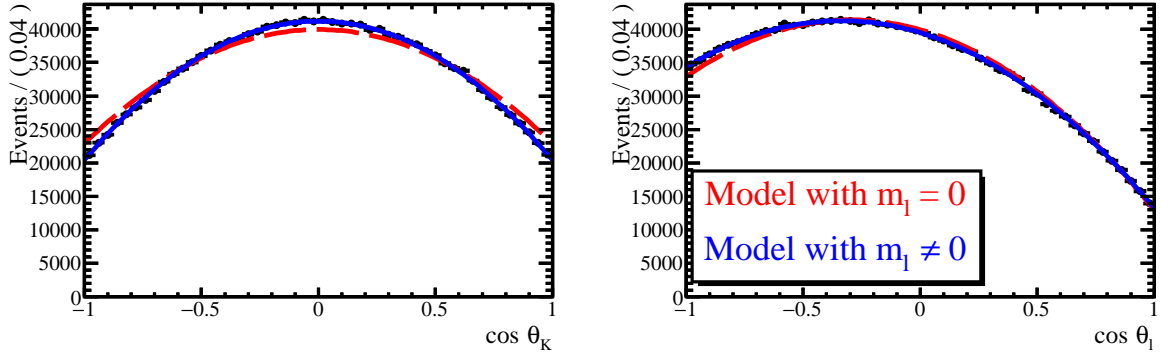}
    \caption{Fit to $1.7\times 10^6$ MC events with 3 components using two models: $m_l$=0(Red) and $m_l\neq0$ (Blue)}
    \label{fig:Fit_to_MC}
\end{figure}

\begin{figure}
    \centering
    \includegraphics[scale=0.80]{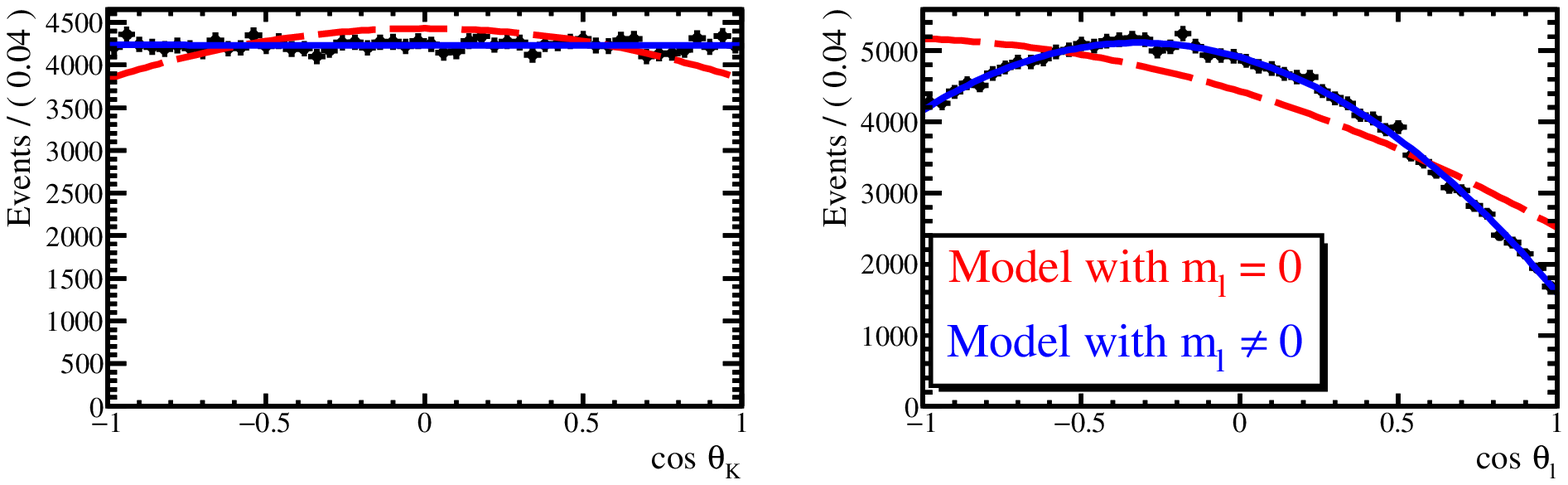}
    \includegraphics[scale=0.80]{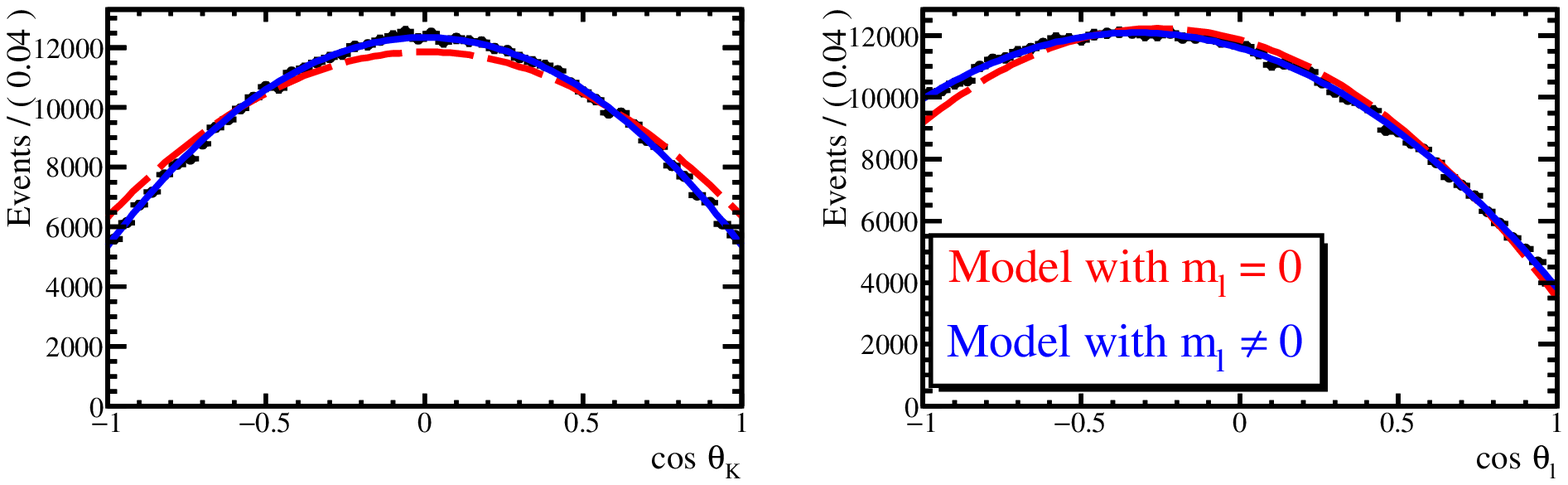}
    \includegraphics[scale=0.80]{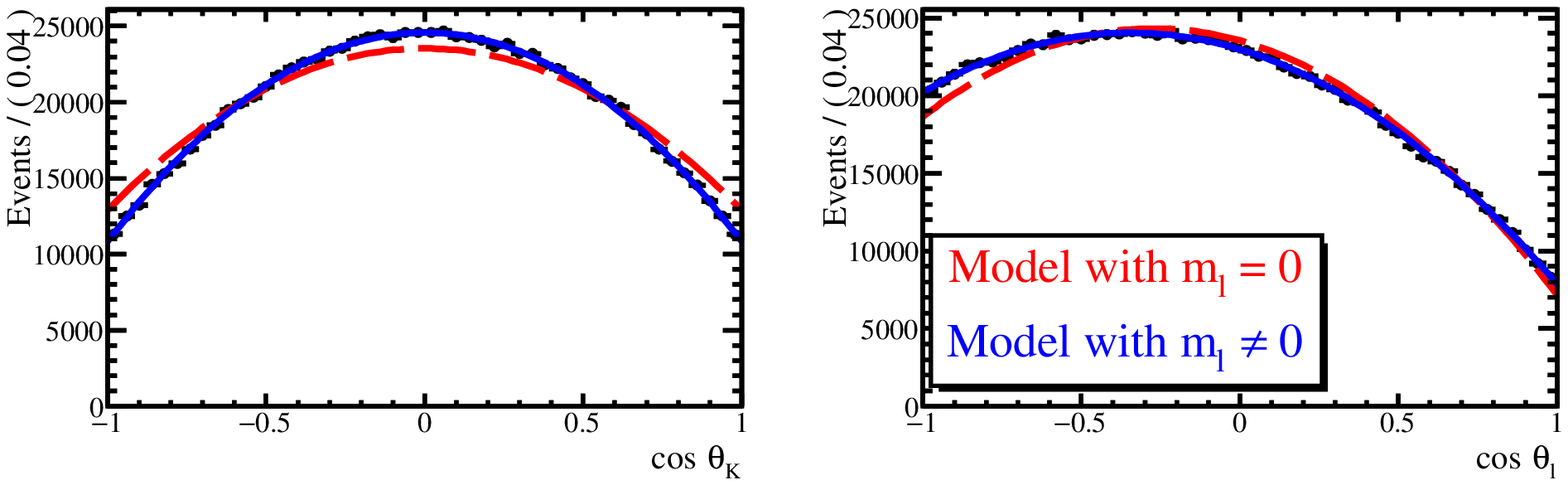}
    \caption{Fit to $0.2\times 10^6$ $D\to K_{1}(\to K\pi\pi)\mu^+\nu_{\mu}$ (upper), $0.5\times 10^6$ $D\to K_{1}(\to K^*(892)(\to K\pi)\pi)\mu^+\nu_{\mu}$ (middle) and $10^6$ $D\to K_{1}(\to \rho(\to \pi\pi)K) \mu^+\nu_{\mu}$(lower)  MC events using two models: $m_l$=0(Red) and $m_l\neq0$ (Blue)}
    \label{fig:Fit_to_MC_0}
\end{figure}

As an illustration, we use a standalone fast simulation software RapidSim~\cite{Cowan:2016tnm} with the LHCb geometrical acceptance to generate MC samples. $D^0\to K_1(1270)^- \mu^+\nu_\mu$ decays are described by EVTGEN~\cite{Lange:2001uf} with the ISGW2 model~\cite{Scora:1995ty}. In the simulation, we have generated about $1.7\times 10^6$ events with three decay modes for $K_1(1270)^-$: about $10^6$ events for $D\to K_{1}(\to \rho(\to \pi\pi)K) \mu^+\nu_{\mu}$,  $0.5\times 10^6$ events for $D\to K_{1}(\to K^*(892)(\to K\pi)\pi)\mu^+\nu_{\mu}$ and $0.2\times 10^6$  for  $D\to K_{1}(\to K\pi\pi)\mu^+\nu_{\mu}$ in which the $K_1$ decay is produced according to phase space.
A comparison of  fitting  these MC samples using the formula~\ref{con:formula_mu} and the one from previous work~\cite{Wang:2019wee} is given in Fig.~\ref{fig:Fit_to_MC}, and   a detailed  analysis of each component is given in Fig.~\ref{fig:Fit_to_MC_0}. Through these figures, one can see that our simulated angular distributions can not be well described by the angular distribution in the previous work~\cite{Wang:2019wee}. 
With the inclusion of the  muon mass, the agreement between theoretical description  and angular distributions of MC events is greatly improved.

\section{Summary}

Weak decays of heavy quarks have played an important role in testing standard model and probing   new physics beyond. Recent studies of flavor-changing neutral current process has revealed some hints for potential NP effects (see for instance Ref.~\cite{Aaij:2021vac}), but a conclusive result is far from well-established, and requests more dedicated theoretical and experimental  studies in future~\cite{Cerri:2018ypt}. At the same time, the photon helicity in $b\to s\gamma$ might   render very competitive potentials for new physics~\cite{Aaij:2020umj}.

Based on  a previous proposal to pin down hadronic uncertainties in $B\to K_1\gamma$~\cite{Wang:2019wee}, 
we  have  in this work systematically  derived differential decay widths and  angular distributions for the decay cascade $D\to K_{1}(1270,1400)\ell^+\nu_{\ell}\to (K\pi\pi)\ell^+\nu_{\ell} (\ell=e,\mu)$. In the derivation,   the mass of electron/muon  is explicitly included. Using the $D\to K_1$ form factors from light-front quark model, we have calculated partial  decay widths and branching fractions for $D^0\to K_1^-\ell^+\nu_{\ell}$ and $D^+\to K_1^0\ell^+\nu_{\ell}$, but pointed out   that these theoretical   results for ${\cal B}(D\to K_1e^+\nu_{e})$ are about a factor 1.5 higher than recent BESIII measurements.

With the angular coefficients, we have demonstrated that the measurement of up-down asymmetry in  $D\to K_{1}e^+\nu_e\to (K\pi\pi)e^+\nu_{e}$ and  the angular distribution in  $D\to K_{1}\ell^+\nu_\ell\to (K\pi\pi)\ell^+\nu_{\ell}$ can help to pin down hadronic uncertainties  in $B\to K_1(\to K\pi\pi)\gamma$. Based on the Monte-Carlo simulation, we have found that after including the muon mass, the angular distributions can be well described by the theoretical framework.

\section*{Acknowledgement}
W.W. thanks Fu-Sheng Yu and Zhen-Xing Zhao for valuable discussions and the collaboration at the early stage of this work.  The authors are grateful to Fei Huang,  Xian-Wei Kang,  Xiao-Rui Lyu, Wen-Bin Qian,  Yang-Heng Zheng for useful discussions.  This work is supported in part by Natural Science Foundation of China under
grant No.11735010, U1932108, U2032102, 12061131006, and 12061141006,  by Natural Science Foundation of Shanghai under grant No. 15DZ2272100.

\end{document}